\documentclass[conference]{IEEEtran}
\IEEEoverridecommandlockouts

\usepackage{amssymb}
\usepackage[cmex10]{amsmath}
\usepackage{stfloats}
\usepackage{graphicx}
\usepackage{subfigure}
\usepackage{tabularx}
\usepackage{epsfig,epsf,color,balance,cite}
\usepackage{verbatim}
\usepackage{url}
\usepackage{bm}
\usepackage{booktabs}
\usepackage{cite}
\usepackage{graphicx}
\usepackage{epstopdf}

\usepackage{algorithm}
\usepackage{algorithmic}

\usepackage{color}
\definecolor{myc1}{rgb}{0,0,0}

\author{
\IEEEauthorblockN{Xingtao Yang\IEEEauthorrefmark{5},
                  Zhenghe Guo\IEEEauthorrefmark{6},
                  Siyun Liang\IEEEauthorrefmark{2},
                  Zhaohui Yang\IEEEauthorrefmark{1}\IEEEauthorrefmark{3},
                  \IEEEauthorrefmark{4},
                  Chen Zhu\IEEEauthorrefmark{2}
                  and Zhaoyang Zhang\IEEEauthorrefmark{1}\IEEEauthorrefmark{3}
                 }
	\IEEEauthorblockA{
			$\IEEEauthorrefmark{1}$College of Information Science and Electronic Engineering, Zhejiang University, Hangzhou, China\\
                $\IEEEauthorrefmark{2}$Polytechnic Institute, Zhejiang University, Hangzhou, Zhejiang, 310015, China\\
			$\IEEEauthorrefmark{3}$Zhejiang Provincial Key Laboratory of Info. Proc., Commun. \& Netw. (IPCAN), Hangzhou, China\\ 
                $\IEEEauthorrefmark{4}$Department of Engineering, King's College London, London, UK\\
                $\IEEEauthorrefmark{5}$ College of Computer Science and Technology, Zhejiang University, Hangzhou, China \\
                $\IEEEauthorrefmark{6}$ Chu Kochen Honors College, Zhejiang University, Hangzhou, China \\
			E-mails: 
   \{ 3230103191, 3230100923, siyunliang, yang\_zhaohui, zhuc,ning\_ming\}@zju.edu.cn
		}}
\begin{document}

% paper title
\title{Rate Maximization for UAV-assisted ISAC System with Fluid Antennas}

% make the title area
\maketitle

\begin{abstract}
This letter investigates the joint sensing problem between unmanned aerial vehicles (UAV) and base stations (BS) in integrated sensing and communication (ISAC) systems with fluid antennas (FA). In this system, the BS enhances its sensing performance through the UAV's perception system. We aim to maximize the communication rate between the BS and UAV while guaranteeing the joint system's sensing capability. By establishing a communication-sensing model with convex optimization properties, we decompose the problem and apply convex optimization to progressively solve key variables. An iterative algorithm employing an alternating optimization approach is subsequently developed to determine the optimal solution, significantly reducing the solution complexity. Simulation results validate the algorithm's effectiveness in balancing system performance.
\end{abstract}

\IEEEpeerreviewmaketitle

\section{Introduction}

As the vision for Sixth Generation (6G) wireless communication systems progressively materializes, the demand for higher data rates, lower latency, broader connectivity, and more diverse application scenarios is becoming increasingly urgent \cite{r3}. Integrated sensing and communication (ISAC), as a revolutionary technology paradigm, achieves deep integration of communication and sensing functionalities by sharing hardware platforms\cite{yang2025privacy}, spectrum resources, and signal processing algorithms\cite{r8}. This significantly enhances spectral efficiency, energy efficiency, and reduces system cost and complexity, thereby establishing ISAC as a core research direction for 6G\cite{r4} \cite{yang2023energy}.

Unmanned aerial vehicles (UAVs), with their unique advantages such as high mobility, flexible deployment, and the ability to readily establish Line-of-Sight (LoS) links, have demonstrated immense potential in both wireless communication and sensing domains\cite{xu2025newpathwayintegratedlearning}. Introducing UAVs into ISAC systems can further extend the coverage of the system, enhance its adaptability in complex environments, and foster various novel applications such as environmental monitoring, intelligent transportation, and emergency rescue. However, efficiently jointly designing communication and sensing functionalities on dynamic UAV platforms and in complex electromagnetic environments still presents numerous challenges\cite{r7}.

To further enhance system performance, research on novel antenna technologies is crucial. Fluid antennas (FA), also known as fluid antenna systems (FAS), are an emerging antenna technology that allows antenna elements to flexibly change their positions and configurations within a predefined space, thereby enabling dynamic reconfiguration of the antenna array\cite{r1}. Compared to traditional fixed-position antennas, FAs can adaptively adjust the activation positions of antenna ports according to channel conditions and service requirements. This allows for superior beamforming gains, interference suppression capabilities, and spatial resolution, offering new opportunities for breakthroughs in ISAC system performance\cite{r10}.

However, applying FAs to UAV-assisted ISAC systems also introduces new dimensions and complexities of the design. Specifically, it is vital to maximize the transmission rate to ensure the timely communication between the UAV and the BS, while also ensuring sufficient power for sensing.

The objective of this paper is to jointly design the beamforming and the selection of activated ports at the UAV, while considering the sensing power constraint at a given direction. The main contributions are as follows.
\begin{itemize}
    \item We investigates an ISAC system assisted by UAVs employing FAs. In this system, the UAV acts as an aerial base station, simultaneously providing communication services to a ground BS and sensing a ground target. Our objective is to maximize the downlink communication rate by jointly optimizing the activated port selection of the FAs on the UAV and the communication beamforming vector, subject to sensing beampattern gain constraints. 
    \item We formulate a joint optimization problem encompassing both communication and sensing models. This problem is typically non-convex and involves the coupled optimization of discrete port selection variables and continuous beamforming variables, making it challenging to solve. To address this, we propose an iterative algorithm based on alternating optimization (AO). In this algorithm, we decompose the original problem into two subproblems: a beamforming optimization subproblem with fixed port selection, and a port selection optimization subproblem with fixed beamforming. For the beamforming optimization subproblem, we use convex optimization to obtain a solution. For the port selection subproblem, we design an efficient search strategy to avoid complex integer programming.
    \item Simulation results validate the effectiveness of the proposed algorithm in balancing communication performance with sensing requirements and demonstrate the potential of FAs in enhancing ISAC system performance.
\end{itemize}

\textit{Notations:} Throughout this paper, we employ standard mathematical notations where $\mathrm{tr}(\cdot)$, $(\cdot)^\mathrm{H}$, and $(\cdot)^\mathrm{T}$ denote the trace, conjugate transpose, and transpose operations respectively; $\Vert \boldsymbol{x}\Vert^2$ represents the 2-norm of vector $\boldsymbol{x}$; $\boldsymbol{W}\succeq \boldsymbol{0}$ indicates that matrix $\boldsymbol{W}$ is positive semidefinite;  $\mathcal{C}\mathcal{N}(0,\sigma^2)$ signifies circularly symmetric complex Gaussian distribution with mean 0 and variance $\sigma^2$.
\section{System Model}
\begin{figure}
    \centering
    \includegraphics[width=0.9\linewidth]{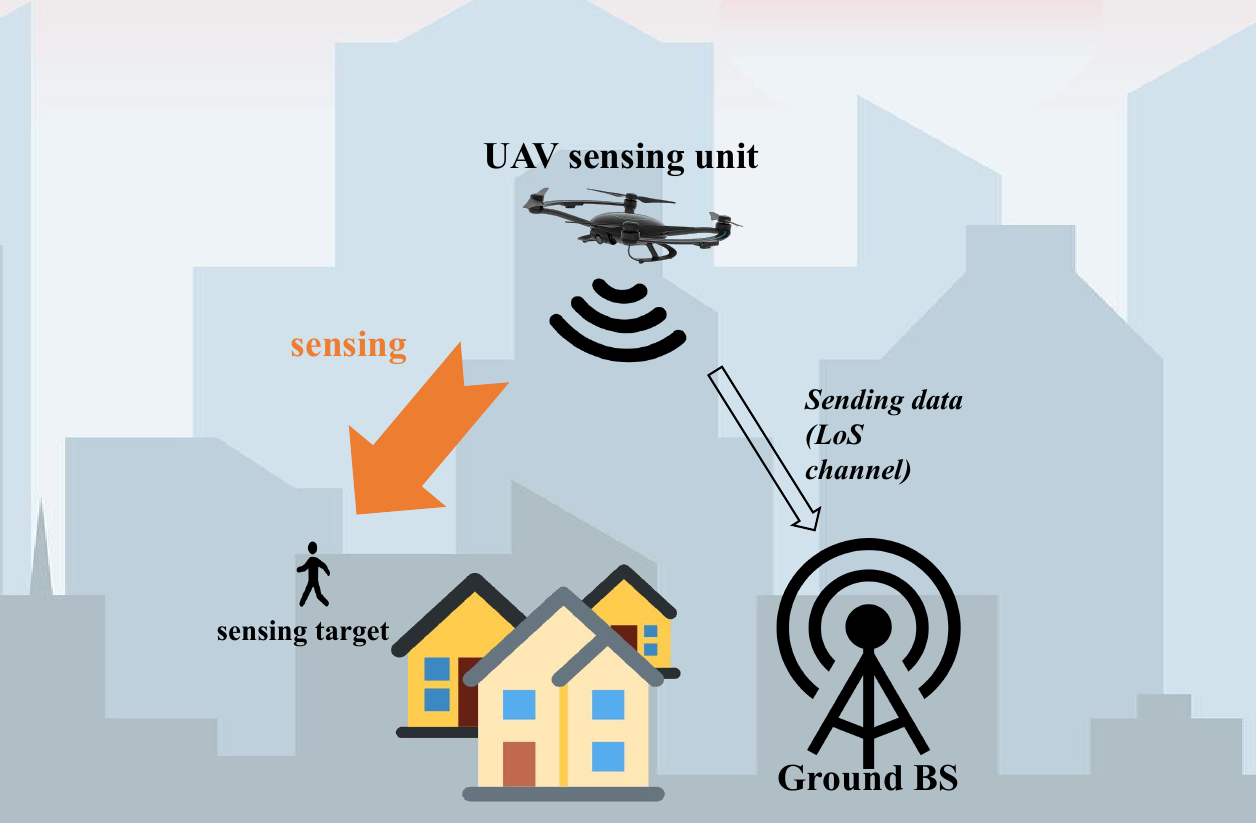}
    \caption{Illustration of the considered ISAC system.}
    \label{fig:system}
\end{figure}

In this paper, a UAV serves as a remote sensing unit, communicating with a ground BS and sensing a ground target. It keeps transmitting sensing data to the ground BS, which is equipped with a uniform linear antenna array with $N$ static antennas. An illustration of the proposed ISAC system is given in Fig.~\ref{fig:system}.

To enhance the communication and sensing performance, the UAV is equipped with fluid antenna system (FAS), with $M$ ports aligned linearly. During operation, it selects $m_0$ activated ports out of the $M$ total ports to send signals.

We assume that the UAV hovers at a given spot $p_U=[x_U,y_U]$, and the communicating BS locates at $p_C=[x_C,y_C]$. For convenience, the power consumption for UAV to hover is denoted by $P_U$. In this paper, we propose there is no natural effect on the UAV, so during the task the UAV stays at the same position, and the power consumption $P_U$ is static.

\subsection{Communication Model}
The beamforming vector of the UAVis denoted by $\boldsymbol{w}$. The sent signal is $x$ with its expectation satisfies $\mathbb{E}\{\vert x\vert^2\}=1$, and the signal received is given by
\begin{equation}
\boldsymbol{y_r}=\boldsymbol{G}^H\boldsymbol{w}\boldsymbol{x}+\boldsymbol{z},
\end{equation}
where $\boldsymbol{z} \in \mathcal{C}^{N\times1}\sim \mathcal{C}\mathcal{N}(0,\sigma^2 \mathrm{\boldsymbol{I}}_{N})$ is the complex additive white Gaussian noise at the communicate BS, and $\boldsymbol{G}$ is the transmit matrix.

Since that the UAV is able to maintain a stationary position in the air, thus enabling direct flight to a location where there is no obstruction between the UAV and the BS, we assume that the UAV-to-ground links are line of sight (LoS) channels. 
For the transmitter on the UAV, $M$ ports of the FAS are aligned evenly perpendicular to the ground, with a distance of $d_{U}$ between two neighbouring ports. The number of activated ports is denoted by $m_0$, which satisfies $1\leq m_0\leq M$. The indices of the selected activated ports are represented as $\boldsymbol{r}=[r_{1}, \cdots, r_{m}, \cdots, r_{m_0}]^T \in \mathbb{Z}^{m_0\times 1}$, where $r_{m} \in \{1,2,\ldots,M\},\;r_{1}<r_{2}<\cdots<r_{m_0}$. We number the ports in a manner that ports with larger indices are more distant from the ground. So the 3D coordinate of the $r_{m}$-th port is 
\begin{equation}
    \boldsymbol{p}_r(m)=[x_U,y_U,H+y_{m}],
\end{equation}
in which $[x_U,y_U]$ is the horizontal position of the UAV, $H$ is the height between the centre of the received antennas and the FA, and $y_{m}$ is given by:
\begin{equation}
    y_{m}=\frac{2(r_{m}-1)-M+1}{2} d_{U}.
\end{equation}

Similarly, we establish a local coordinate at the ground BS side, and the y-coordinate of the $n$-th received antenna is given by 
\begin{equation}
    y^r_{n}=\frac{2(n-1)-N+1}{2} d_{C}.
\end{equation}
where $d_{C}$ denotes the distance between adjacent received antennas.

To construct the transmit matrix $\mathbf{G}$, we need to obtain the steering vector. When communicating to BS, the propagation path distance difference to the $n$-th received antenna for the $m$-th port between it and the centre of FAS is 
\begin{equation}\label{eq:dc}
    d_C(m,n)= \sqrt{L_C^2+(H+y_m-y^r_n)^2}-\sqrt{L_C^2+H^2}.
\end{equation}
where $L_C=\Vert p_C-p_U\Vert$ is the horizontal distance between the UAV and the BS.

When the propagation path distance difference \eqref{eq:dc} is known,  it is possible to derive the phase differences of transmit antennas. With the wavelength $\lambda$, the transmit response vector for the $n$-th received antenna is
\begin{equation}
    \boldsymbol{g}_C(n)\triangleq\left[e^{j\frac{2\pi}{\lambda}d_C(1,n)}, \ldots,e^{j\frac{2\pi}{\lambda}d_C(m_0.n)}\right]^T\in \mathbb{C}^{m_0\times1}.
\end{equation}

Furthermore, we can obtain the response matrix $\boldsymbol{G}$:
\begin{equation}\label{eq:G}
    \boldsymbol{G}=[\boldsymbol{g}_C(1),\ldots\boldsymbol{g}_C(N)]\in \mathcal{C}^{m_0\times N}.
\end{equation}

Based on \eqref{eq:G}, the achievable communication rate is given by
\begin{equation}
    R=\log \det \left(\mathbf{I}_N+\frac{ \boldsymbol{G}^\mathrm{H}\boldsymbol{W}\boldsymbol{G}}{\sigma^2}\right),
\end{equation}
where $\boldsymbol{W}$ denotes the transmit covariance matrix of the FAS loaded on the UAV.
\subsection{Sensing Model}
In this paper, we assume that the UAV keeps sensing at a static target. The steering vector for sensing is given as 
\begin{align}\label{eq:at}
    \boldsymbol{a}(\theta) &= 
    \left[1, e^{j\frac{2\pi}{\lambda}d_U(r_m-r_1)\sin\theta}, \ldots,e^{j\frac{2\pi}{\lambda}d_U(r_{m_0}-r_1)\sin\theta} \right]^T \\ \nonumber
    & \quad \in \mathbb{C}^{m_0\times1}.
\end{align}
where $\theta$ represents the angle for sensing direction.

With \eqref{eq:at}, the sensing beampattern gain is given as
\begin{equation}
    P_S(\theta)=\boldsymbol{a}(\theta)^H\boldsymbol{W}\boldsymbol{a}(\theta).
\end{equation}

To ensure sensing performance, $P_S$ should follows
\begin{equation}
    P_S\geq \Gamma,
\end{equation}
where $\Gamma$ is the predefined sensing beampattern gain. 
\subsection{Problem Formulation}
In this paper, we aim at maximizing the transmission rate while having satisfying sensing performance. The maximization problem is formulated as follows:
    \begin{align}
        \max_{\boldsymbol{W},\boldsymbol{r}} \quad& R,
         \label{eq:mp}\\
    \textrm{s.t.}\quad 
    &r_{m}\in \{1,\ldots,M\},\tag{\ref{eq:mp}{a}} \label{c1}\\
    &r_{1}< r_{2}< \cdots < r_{m_0},\tag{\ref{eq:mp}{b}} \label{c2}\\
    &\mathrm{tr} (\boldsymbol{W})+P_U \leq P_{\max}, \tag{\ref{eq:mp}{c}} \label{c3}\\
    &P_S(\theta)\geq \Gamma \tag{\ref{eq:mp}{d}}\label{c4}\\
    &\boldsymbol{W}\succeq \boldsymbol{0}, \tag{\ref{eq:mp}{e}}\label{c5}
    \end{align}

Constraint \eqref{c1} and \eqref{c2} are the numerical constraints for port indices in FAS. Constraint \eqref{c3} forbids exceeding total power consumption, and constraint \eqref{c4} ensures the sensing power is larger than the pre-defined threshold $\Gamma$. Lastly, constraint \eqref{c5} holds the nature of transmit covariance matrix $\boldsymbol{W}$.
\section{Proposed Algorithm}
Due to the discrete nature of port selection vector $\boldsymbol{r}$, the problem is non-convex and hard to solve directly. So we propose an alternative optimization algorithm, divide problem \eqref{eq:mp} into 2 subproblems and solve them alternatively, until the solution converges at given accuracy.
\subsection{Optimization for Transmit Covariance Matrix $\boldsymbol{W}$}
We first consider optimize the transmit covariance matrix $\boldsymbol{W}$.  With fixed port selection $\boldsymbol{r}$, problem \eqref{eq:mp} is reduced to 
    \begin{align}
        \max_{\boldsymbol{W}} \quad& R,
         \label{eq:W}\\
    \textrm{s.t.}\quad 
    &\mathrm{tr}(\boldsymbol{W})+P_U\leq P_{\max}, \tag{\ref{eq:W}a} \label{eq:Wa} \\ 
    &P_S(\theta)\geq \Gamma \tag{\ref{eq:W}b}\label{eq:Wb} \\
    &\boldsymbol{W}\succeq \boldsymbol{0}.
\end{align}

Because that the hovering power is static, so, by letting $P_C=P_{\max}-P_U$, constraint (13a) is reduced to 
\begin{equation}
    \mathrm{tr}(\boldsymbol{W})\leq P_C.
\end{equation}

Now, problem \eqref{eq:W} is reduced to a convex problem and can be efficiently solved by existing maths tools such as CVX \cite{r5}. 

\subsection{Optimization for Port Selection $\boldsymbol{r}$}
In this subproblem, we focus on optimizing the port selection vector $\boldsymbol{r}$. With a determined $\boldsymbol{W}$, problem \eqref{eq:mp} is reduced to 
 \begin{align}
        \max_{\boldsymbol{r}} \quad& R,
         \label{eq:pr}\\
    \textrm{s.t.}\quad 
    &r_{m}\in \{1,\ldots,M\},\tag{\ref{eq:pr}{a}} \\
    &r_{1}< r_{2}< \cdots < r_{m_0},\tag{\ref{eq:pr}{b}}\\
    &P_S(\theta)\geq \Gamma. \tag{\ref{eq:pr}{c}}
\end{align}
    
The discrete value of $r_m$ makes problem \eqref{eq:pr} hard to solve directly. It is possible to leverage maths tools with integer programming, such as Gurobi, to obtain a solution, but the time complexity will be large as the scale of problem grows. To avoid complex combinations, we seek to optimize $r_m$ instead. In this way, we could formulate a subproblem for optimizing $r_m$, as
\begin{align}
    \max_{r_m} \quad& \log \det \left(\mathbf{I}_N+\frac{ \boldsymbol{G}^\mathrm{H}\boldsymbol{W}\boldsymbol{G}}{\sigma^2}\right),\label{eq:sub}\\
    \textrm{s.t.}\quad 
    &r_{m}\in \{1,\ldots,M\},\tag{\ref{eq:sub}{a}} \label{eq:rm1}\\
    &r_{1}< r_{2}< \cdots < r_{m_0},\tag{\ref{eq:sub}{b}}\label{eq:rm2}\\
    &P_S(\theta)\geq \Gamma. \tag{\ref{eq:sub}{c}} \label{eq:pv}
\end{align}

So, by trying the possible value of $r_m$ according to the constraints \eqref{eq:rm1} and \eqref{eq:rm2}, we can compute the value of \eqref{eq:sub}, and update $r_{m}$ when the new value of \eqref{eq:sub} is larger than the current one and the new $r_{m}$ does not violate \eqref{eq:pr}, which ensures that the value of \eqref{eq:sub} is non-strictly increasing within the optimization process.

The process of the overall algorithm is as follows:
\begin{algorithm}
    \caption{Alternate optimization for beamforming matrix and port selection}
    \begin{algorithmic}\label{al:2}
        \STATE \textbf{Initialize:} Iteration index $i=0$, accuracy $\epsilon$, $\boldsymbol{r}^{(0)}$.
        \REPEAT
        \STATE Obtain $\boldsymbol{W}$ from solving \eqref{eq:W}.
        \FOR{$m$=1 to $m_0$}
            \STATE Find the $r_m$ which maximizes \eqref{eq:sub}
        \ENDFOR
        \STATE Set $i=i+1$
        \STATE Calculate $R^{(i)}(\boldsymbol{W}, \boldsymbol{r})$
        \UNTIL{$|R^{(i)}(\boldsymbol{W}, \boldsymbol{r})-R^{(i-1)}(\boldsymbol{W}, \boldsymbol{r})|\leq \epsilon$}
    \end{algorithmic}
\end{algorithm}

Sub-problem (13) is solved by convex problem solvers has promising convergence, while sub-problem (15) is solved by an exhaustive-search-like method. The value of objective function (12) keeps increasing during the optimization and as (12) is boundary, the overall algorithm is promised to converge.
\section{Simulation Results}
All the simulation parameters are summarized in Table~\ref{tab:my_label}. We focus on the convergence behavior and the final performance in terms of communication rate and port selection.
\begin{table}[!h]
    \caption{Simulation Parameters}
    \centering
    \begin{tabular}{|c|c|c|}
        \hline
        \textbf{Notation} & \textbf{Parameter} & \textbf{Value}\\
        \hline
        $M$ & Total number of FAS ports	 & 40\\
        \hline
        $m_0$ & Number of ports to communicate & 10\\
        \hline
       $ P_{max}$ & Maximum transmit power & 10 dBm\\
        \hline
        $P_U$ & UAV internal power consumption & 7 dBm\\
        \hline
        $\Gamma$ & Perception direction threshold	& 8 dBm\\
        \hline
        $\theta$ & Perception direction angle & $\pi / 6$\\
        \hline
        $\lambda$ & Wavelength & 0.1 m\\
        \hline
        $d$ & Port spacing & 0.05 m\\ 
        \hline
        $H$ & UAV altitude & 20 m\\
        \hline
        $\epsilon$ & Convergence threshold & 0.001\\
        \hline 
    \end{tabular}
    \label{tab:my_label}
\end{table}

\begin{figure}
    \centering
    \includegraphics[width=\linewidth]{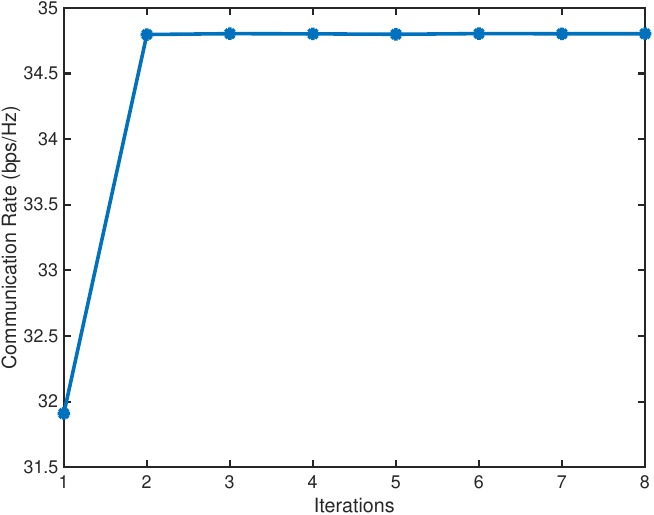}
    \caption{Convergence Behavior of Communication Rate}
    \label{fig:diagram1}
\end{figure}

Fig.~\ref{fig:diagram1} illustrates the evolution of the communication rate with respect to the iteration index during the optimization process.It can be observed that the communication rate $R$ increases rapidly in the initial iterations.Furthermore, it is shown that, although slight fluctuations may occur during the iterative process, the final result remains stable and satisfies the sensing constraint. This implies that the proposed method can dynamically balance the trade-off between communication performance and sensing requirements by jointly adapting the resource allocation. The convergence of the port selection vector also reflects the algorithm’s capability in finding a near-optimal solution over a discrete feasible set, which is typically difficult to optimize directly.

\begin{figure}
    \centering
    \includegraphics[width=\linewidth]{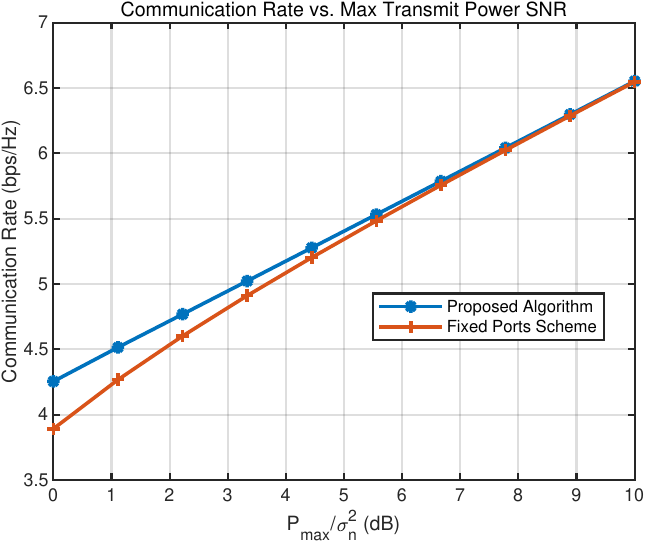}
    \caption{Communication Rate vs. Max Transmit Power SNR}
    \label{fig:diagram2}
\end{figure}

In Fig.~\ref{fig:diagram2}, the simulation results illustrate the communication rate performance of the proposed algorithm compared to a fixed port baseline at various maximum transmit power levels $P_{max}$. Although the rates of both schemes are monotonically increasing functions of $P_{max}$, the advantage of the proposed algorithm is most pronounced in the power-limited regime. As the transmit power becomes abundant, the performance bottleneck shifts away from the power itself and the rates of the two schemes converge. This confirms that the algorithm's core contribution is its efficient power allocation, which maximizes the communication rate under strict power budgets. This capability is of significant theoretical and practical importance for energy-constrained and power-sensitive wireless applications.

\begin{figure}
    \centering
    \includegraphics[width=\linewidth]{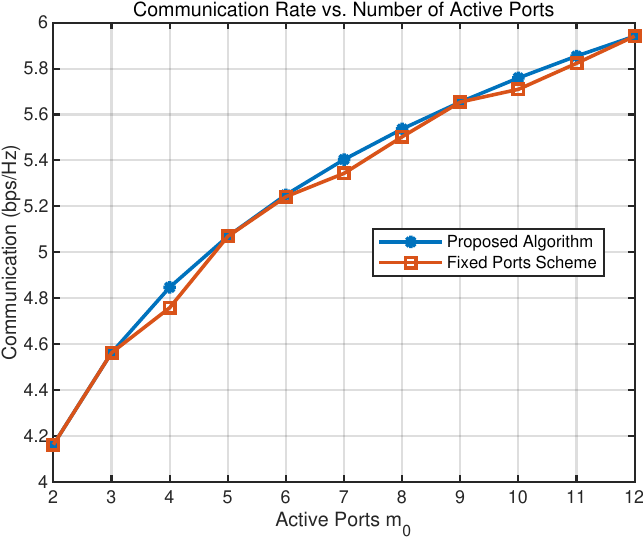}
    \caption{Communication Rate vs. Number of Active Ports}
    \label{fig:diagram3}
\end{figure}

Fig.~\ref{fig:diagram3} reveals the relationship between the communication rate and the number of active ports, $m_0$. In general, the communication rate is a monotonically increasing function of $m_0$, which verifies the positive effect of increasing the aperture of the antenna array to improve the beamforming gain.

It should be noted that the performance gain offered by the proposed optimization algorithm exhibits an intermittent characteristic. For certain values of $m_0$, the performance of the algorithm perfectly coincides with that of the fixed port scheme, suggesting that conventional port selection is already a locally optimal solution under these specific conditions. However, for numerous other values, the optimization algorithm achieves significant rate enhancements, outperforming conventional port selection in the majority of cases. This provides strong evidence for the algorithm's effectiveness and robustness: it consistently matches or exceeds the performance of the baseline scheme by leveraging an intelligent port selection strategy to exploit additional performance potential as channel conditions allow.

In general, the proposed iterative framework demonstrates reliable convergence properties and strong adaptability, making it suitable for practical ISAC applications with complex design requirements.

\section{Conclusions}
We investigated a UAV-assisted ISAC system with joint design between the UAV and the base station, aiming to maximize the system’s communication performance while ensuring reliable sensing capabilities. To address the initially complex problem, we decomposed it into more manageable subproblems and adopted an alternating optimization framework to reduce computational complexity. Techniques such as convex optimization were employed to obtain an efficient solution. Simulation results validated the effectiveness of the proposed algorithm. This approach is expected to be applicable to future ISAC systems where communication performance is prioritized.
\bibliographystyle{IEEEtran}
\bibliography{main}
\end{document}